\documentclass[acmsmall]{acmart}

\AtBeginDocument{%
  \providecommand\BibTeX{{%
    \normalfont B\kern-0.5em{\scshape i\kern-0.25em b}\kern-0.8em\TeX}}}

\acmBooktitle{Woodstock '18: ACM Symposium on Neural Gaze Detection,
  June 03--05, 2018, Woodstock, NY}

\usepackage{graphicx}
\usepackage{soul}
\usepackage{xcolor}
\usepackage{amsmath}

\usepackage{amssymb}
\usepackage{multirow}
\usepackage{booktabs}
\usepackage{arydshln}

\begin{document}

\title{MetaDetector: Meta Event Knowledge Transfer for Fake News Detection}

\author{Yasan Ding}
\affiliation{%
  \institution{Northwestern Polytechnical University}
  \city{Xi'an}
  \country{P.R.China}}
\email{yasanding@mail.nwpu.edu.cn}

\author{Bin Guo}
\authornote{Corresponding author}
\email{guob@nwpu.edu.cn}
\affiliation{%
  \institution{Northwestern Polytechnical University}
  \city{Xi'an}
  \country{P.R.China}
}

\author{Yan Liu}
\affiliation{%
  \institution{Northwestern Polytechnical University}
  \city{Xi'an}
  \country{P.R.China}
}
\email{yan_emily@outlook.com}

\author{Yunji Liang}
\affiliation{%
 \institution{Northwestern Polytechnical University}
 \city{Xi'an}
 \country{P.R.China}}
\email{liangyunji@nwpu.edu.cn}

\author{Haocheng Shen}
\affiliation{%
  \institution{Northwestern Polytechnical University}
  \city{Xi'an}
  \country{P.R.China}}
\email{947734694@qq.com}

\author{Zhiwen Yu}
\affiliation{%
  \institution{Northwestern Polytechnical University}
  \city{Xi'an}
  \country{P.R.China}}
\email{zhiwenyu@nwpu.edu.cn}

\renewcommand{\shortauthors}{Yasan Ding, et al.}

\begin{abstract}
  The blooming of fake news on social networks has devastating impacts on society, economy, and public security. Although numerous studies are conducted for the automatic detection of fake news, the majority tend to utilize deep neural networks to learn event-specific features for superior detection performance on specific datasets. However, the trained models heavily rely on the training datasets and are infeasible to apply to upcoming events due to the discrepancy between event distributions. Inspired by domain adaptation theories, we propose an end-to-end adversarial adaptation network, dubbed as \textit{MetaDetector}, to transfer meta knowledge (event-shared features) between different events. Specifically, \textit{MetaDetector} pushes the feature extractor and event discriminator to eliminate event-specific features and preserve required event-shared features by adversarial training. Furthermore, the pseudo-event discriminator is utilized to evaluate the importance of historical event posts to obtain partial shared features that are discriminative for detecting fake news. Under the coordinated optimization among the four submodules, \textit{MetaDetector} accurately transfers the meta-knowledge of historical events to the upcoming event for fact checking. We conduct extensive experiments on two large-scale datasets collected from Weibo and Twitter. The experimental results demonstrate that \textit{MetaDetector} outperforms the state-of-the-art methods, especially when the distribution shift between events is significant. Furthermore, we find that \textit{MetaDetector} is able to learn the event-shared features, and alleviate the negative transfer caused by the large distribution shift between events.
\end{abstract}



\maketitle

\section{Introduction}
\label{Introduction}
The enthusiasm about social media not only boosts the exchange of information, but also provides ideal platforms for the wide spread of false information, commonly known as \textit{fake news}. Especially in major public events (e.g.,the political election \cite{bovet2019influence} and plague prevention \cite{gallotti2020assessing}), the prevalence of fake news distracts decision makers' attention, causes cognitive misperception among audience, and spreads panic to the public \cite{qiu2017limited}. The mainstream social media platforms have witnessed the outbreak of the \textit{`infodemic'} \cite{cinelli2020covid} about the COVID-19 pandemic. For example, a Facebook post claiming that microwaves could sterilize the used masks has been shared more than 7,000 times and misled a large number of people\footnote{https://www.kiro7.com/news/trending/coronavirus-fact-check-is-microwaving-fabric-mask-good-way-sanitize-it/VVDB4NUKFNE5ZBZLPIVUEDAHME/}. What's worse, nearly 50 Twitter accounts and 400 Facebook communities promoted the conspiracy theory that 5G wireless technology could spread the coronavirus, leading to the destruction of most base stations in Britain\footnote{https://www.nytimes.com/2020/04/10/technology/coronavirus-5g-uk.html}. 

As fake news is flooding social networks, a large number of institutions and researchers are taking actions to curb the dissemination of falsehoods. For instance, \textit{Facebook} and \textit{Twitter} encourage users to flag inaccurate or incorrect posts to remind other potential audiences \cite{10.1007/s11280-012-0188-y}. Several fact-checking websites (e.g., \textit{Politifact.com}\footnote{https://www.politifact.com/} and \textit{Snopes.com}\footnote{https://www.snopes.com/fact-check/}) regularly released evidences (usually provided by experienced officers and journalists) of recently checked fake news. Although the aforementioned evidence-based detection methods are highly interpretable, they are time-consuming and labour-intensive. Fortunately, the group behaviors, social interactions, and community dynamics triggered by news dissemination 
imply the credibility of contents \cite{zhang2011emergence}, which may embrace potential characteristics of real and fake news. Therefore, automatic fake news detection method based on mining content or social context features has become a research hotspot \cite{shu2017fake}. Specifically, content-based methods usually extract lexical, syntactic or topic features of event-related posts to train bi-classifiers for fake news detection \cite{castillo2011information,rashkin2017truth,che2018fake,maddock2015characterizing}; While social context-based methods decide whether a piece of news is real based on its propagation patterns \cite{jin2014news,bian2020rumor}, social interactions \cite{jin2016news,zhao2015enquiring}, and the corresponding disseminators' credibility \cite{shu2018understanding,shu2019beyond}. 

\begin{figure}
\includegraphics[width=\textwidth]{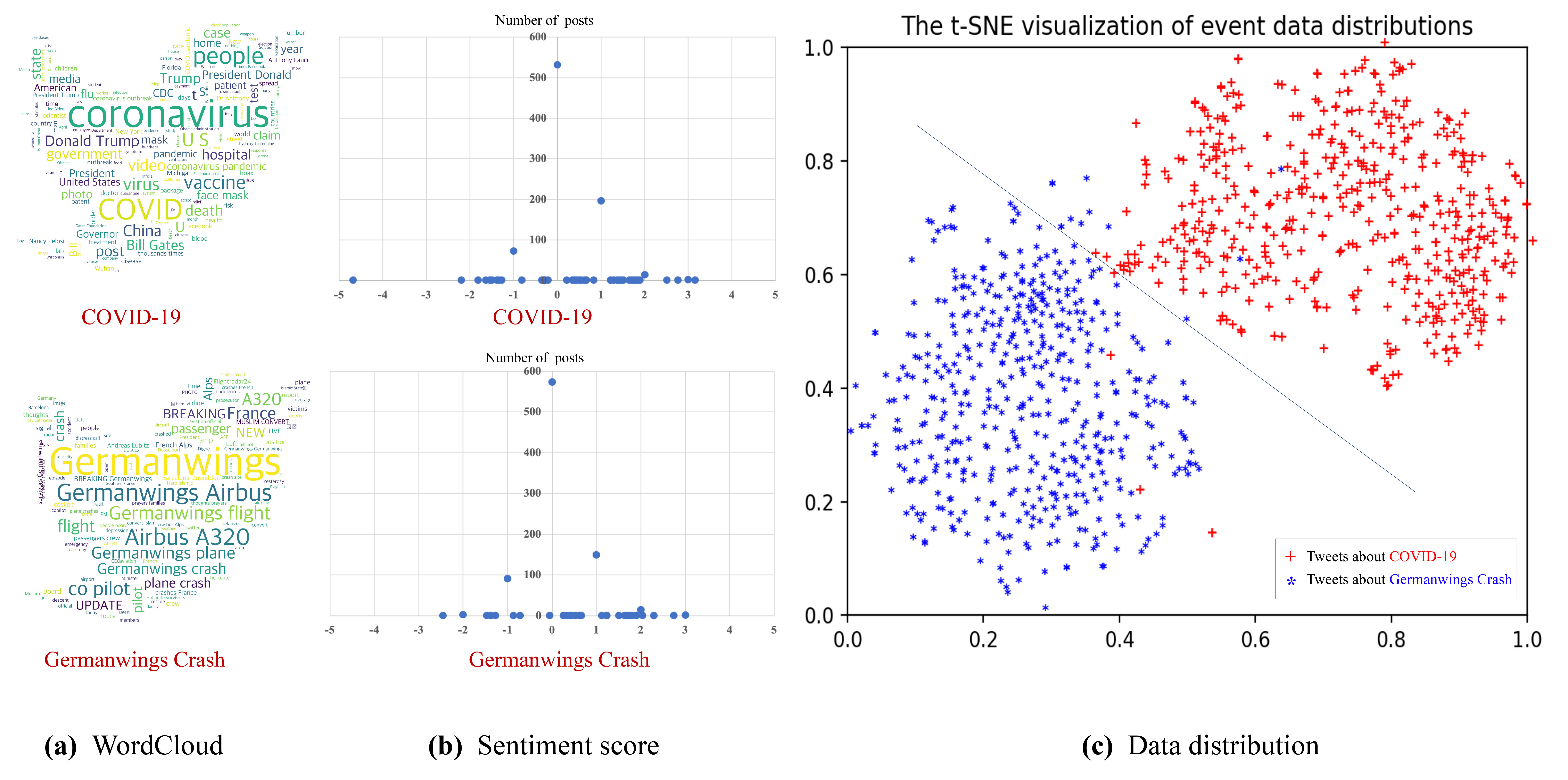}
\caption{An empirical comparison of tweets about the \textit{COVID-19} and the \textit{Germanwings Crash}. From left to right are the wordcloud, sentiment score, and data distribution of sampled tweets from the two events. In the wordcloud, the top figure shows hot words related to COVID-19, and the bottom one belongs to Germanwings Crash. In the sentiment score, we set zero, positive, and negative numbers to indicate neutral, positive, and negative sentiment tendencies. A score with a larger absolute value indicates a stronger intensity, as shown on the horizontal axis of the figures. In the data distribution, we use the 768-dimensional word vectors pre-trained by BERT to represent all the sampled tweets, and visualize their distributions in a two-dimensional plane (red nodes are tweets from COVID-19, and blue nodes are tweets from Germanwings Crash), showing two obvious clusters.} 
\label{Example}
\end{figure}

Although automatic fake news detection is not a new phenomenon, existing methods are still powerless to solve the practical detecting problem due to the lack of \textit{model adaptivity to the new events} \cite{guo2020future}. Most detection approaches use deep neural networks to embed news contents of different events to a high-dimensional feature space for learning latent representations of news. However, the single fake news classification loss induces them to capture event-specific features, which are challenging to be shared with other events. Unfortunately, the data distributions of different news events often deviate from each other, and features change dramatically across new and historical events consequently. As shown in Figure \ref{Example}, we randomly sample 800 tweets about the \textit{COVID-19}\footnote{https://twitter.com/search?q=COVID-19} and the \textit{Germanwings Crash}\footnote{https://twitter.com/search?q=Germanwings\%20crash} respectively on Twitter to empirically analyze their characteristics underlying corresponding keywords, sentiments, and data distributions. The wordcloud illustrates that the tweets in each event have abundant event-specific keywords. For example, \textit{coronavirus}, \textit{vaccine}, \textit{Germanwings}, and \textit{Airbus} etc. only appear in one of the events. While the sentiment sub-figures reveal that most tweets' sentiment values of the two events are in the neutral (equal to `0') and slightly positive (less than `3') range. In addition, we use t-SNE \cite{maaten2008visualizing} to visualize the word embeddings of all the sampled tweets pre-trained by BERT \cite{devlin2018bert} (blue nodes represent tweets about COVID-19, and red nodes indicate tweets about Germanwings Crash in Figure \ref{Example}), which is a strong indicator of the distribution shift between different events. However, existing methods do not take into consideration the event distribution property and only focus on the event-specific features for fake news detection, which are infeasible to discriminate news that have not been seen in training datasets \cite{zhou2020survey}. The development of transfer learning \cite{pan2009survey} provides a feasible solution to this problem. Although we could hardly collect abundant epidemiological data in a limited time for detecting fake news about COVID-19, we have accumulated a large amount of data on other verified events. Transferring the knowledge learned from historical data (hereinafter referred to as ``\textit{source event}'') to the identification of latest news on COVID-19 (hereinafter referred to as ``\textit{target event}'') cultivates to construct a generalized detection model in specific events.

The basis of event knowledge transfer lies in that different news events consist of both \textit{event-specific} information and \textit{event-shared} information. Obviously, social posts are meaningful contents created and manipulated by users based on their own perceptions of news events, which inevitably contain \textit{social clues about target incidents}, \textit{descriptions about the cause of events}, \textit{profiles of some mentioned persons} and other event-specific information. In addition, there are also common or similar information that can appear in multiple events, e.g., \textit{peculiar emotional words}, \textit{psychological signals}, \textit{user stances}, etc. Event knowledge transfer is to transfer event-shared features to guide the fake news detection in upcoming events. For example, Wang et al. \cite{wang2018eann} proposed the \textit{EANN} model to reduce event-specific features and utilized event-shared features to detect fake news in new events. It employed an event discriminator to measure the  dissimilarities between events, reduced the shift in feature distributions of different events by adversarial learning, and finally disclosed transferable features for detecting fake news in the target event. However, \textit{EANN} ignores the importance of each post to the target event. Due to users’ publishing behaviors, the quality of social posts is uneven, resulting in different importance and contribution of each post to new events. It seems to be invalid to those fake news detection methods that apply the attention mechanism to find useful temporal linguistic features \cite{chen2018call,guo2018rumor}. Consequently, it is believed that the same importance of source event posts cannot effectively represent the shared feature space of events. If the discrepancy between events’ feature distributions is noticeable, event-level knowledge transfer may lead to negative transfer, demonstrating that crucial transferable features need to be disentangled. 

In this paper, we refer to crucial event-shared (or transferable) features as \textit{meta knowledge}, and focus on detecting fake news in new events by transferring the meta knowledge learned from historical events. Acquiring meta knowledge is essentially decoupling event-shared and event-specific features from high-level feature space of different events, and learning latent representations of the target event under shared features for fact checking. The primary challenges are shown below:
\begin{itemize}
    \item \textbf{How to distinguish the event-specific features of news posts and reduce them as much as possible}? A straightforward idea is to measure the dissimilarity of content features of different events. The more dissimilar the feature distributions compared, the more likely that the extracted features belong to a specific event. However, the high-dimensional content features learned by deep neural networks are difficult to describe the dissimilarity among events by traditional metrics. During model training, feature embeddings of source and target events are constantly changing. Whether the detection model can capture these changes is vital to the observation of event-specific features.
    \item \textbf{How to manifest which labeled historical post is superior to identifying fake news in unlabeled new events}? Since each post has a different degree of contribution to new event fact checking, it is appropriate to implement post-level knowledge transfer through a weighted mechanism. Unfortunately, the detection model hardly know which historical posts are so relevant and important to upcoming events that they should be introduced into new fake news detection (given higher weights), because all the target posts are unlabeled. If the importance of transferred knowledge cannot be reasonably estimated, it may also cause negative transfer \cite{long2018conditional,zhang2018importance}.
\end{itemize}

To address the above challenges, we propose an end-to-end debunking framework based on adversarial domain adaptation methods, namely \textit{MetaDetector}, which automatically transfers learned meta knowledge from verified events to guide the fake news detection in target events. Our model is mainly composed of the feature extractor, the fake news detector, the event discriminator, and the pseudo-event discriminator (as presented in Figure \ref{MetaDetector}). In response to the first challenge, \textit{MetaDetector} trains feature extractor and event discriminator in a min-max game to gradually reduce event-specific features and retain event-shared features. When features learned by feature extractor are enough to deceive event discriminator from distinguishing the source of them, then transferable features can be extracted. For the second challenge, we incorporate a pseudo-event discrimination-based weighting mechanism to calculate the importance of source event instances. It is based on the observation that the event discriminator indicates probabilities of input features from the source event. The higher the probability score, the more likely the feature is to be derived from the source-event feature space, and it should be given a small weight. Obviously, the transferability of each post can be measured by this indirect weighting mechanism. In order to calculate weights while extracting event-shared features, \textit{MetaDetector} introduces the pseudo-event discriminator, similar to event discriminator, to apply weights to source event posts. Then \textit{MetaDetector} utilizes weighted source posts and unlabeled target posts to corporately train feature extractor and event discriminator for meta knowledge. Finally, fake news detector identifies fake news in the target event based on learned meta knowledge. In summary, the main contributions of our work are as follows:

\begin{itemize}
    \item We explore a practical fake news detection problem of \textit{detecting fake news in upcoming events} from the perspective of knowledge transfer.
    \item We propose a pseudo-event discrimination-based weighting mechanism to fulfill post-level knowledge transfer, which characterizes the relationship between historical events and upcoming events in a fine-grained way.
    \item We propose a general framework for fake news detection in new events, namely \textit{MetaDetector}, which reduces the distribution discrepancy between events, learns the meta knowledge underlying different events, and improves the generalization performance on unseen events by adversarial adaptations networks.
    \item Our experiments on two public authoritative datasets demonstrate the effectiveness of \textit{MetaDetector} in detecting fake news of new events. In addition, we conduct another test on a COVID-19 related fake news dataset\footnote{https://github.com/cyang03/CHECKED}, and analyze the experimental results in detail. 
\end{itemize}

The rest of this paper is organized as follows: we briefly review representative works in Section \ref{Related_Work}, and introduce the details of our proposed \textit{MetaDetector} in Section \ref{Methodology}. We conduct intensive experiments, and the experimental results are shown in Section \ref{Experiments}. Finally, we conclude our work in Section \ref{Conclusion}.
\section{Related Work}
\label{Related_Work}
In this section, we present one brief review about the detection of fake news. Meanwhile, we present the preliminary theory of domain adaptation methods. 
\subsection{Fake News Detection}
The utilization of deep learning methods in fake news detection ameliorates the detection efficiency and accuracy, which can automatically extract the latent high-level feature representations of news events. In fact, there are several tasks, e.g., rumor / misinformation / disinformation detection. This paper adopts a broad definition of fake news, i.e., ``\textbf{fake news is false news, where news broadly includes claims, statements, posts, among other types of information}'' \cite{zhou2020survey}. Subsequently, we review prior works in the following three categories: content-based, and social context-based detection.

Content-based detection methods mainly construct the whole event posts into time-series segments, and then feed them into deep neural networks (DNNs) to extract latent semantic features of news events for detection. For example, Chen et al. \cite{chen2018call} and Ma et al. \cite{ma2016detecting} divide tweets per event into fixed-length and variable-length time series respectively, and feed them into recurrent neural networks (RNN) and their variants to capture temporal-linguistic features for fake news detection. In addition, Yu et al. \cite{yu2017convolutional} state that RNN-based methods are potentially biased against the latest input posts and inadequate to detect fake news in advance. Therefore, they propose a CNN-based detection model named \textit{CAMI}, which flexibly extracts key features scattered in related posts through convolution operations and shapes the interaction among high-level features.

Social context-based detection methods principally embed social interactions (e.g., user comments/likes/reposts) or information diffusion structures into dense vectors by neural networks for following detection. For instance, Liu et al. \cite{liu2018early} observe that real and fake news have different disseminated patterns, so they use gated recurrent unit (GRU) and CNN to extract global and local features of the retweeting sequences for fake news detection. On this basis, Lu et al. \cite{lu2020gcan} introduce the graph convolution neural networks (GCNs) to learn more accurate structural information of news propagation paths.

Different from previous works, Bian et al. \cite{bian2020rumor} emphasize the wide dispersion of fake news in addition to its deep propagation, and consequently propose the \textit{Bi-GCN} model to comprehensively describe the spread of fake news in social networks. Furthermore, there are also several works comprehensively utilize content features and social context features (news articles, user profiles, and social interactions) \cite{shu2019beyond}, such as CSI \cite{ruchansky2017csi}, dEFEND \cite{shu2019defend}, and FANG \cite{nguyen2020fang}. Existing methods tend to improve fake news detection metrics on specific datasets. However, the prior studies heavily rely on the training datasets and are infeasible to apply to unseen events due to the discrepancy between event distributions. How to use the knowledge learned from historical events to detect fake news in upcoming events has not been studied yet.
\subsection{Domain Adaptation}
Domain adaptation mainly tackles the problem of knowledge transfer where the source domain and the target domain have different marginal probability distributions but the same conditional probability distributions \cite{zhang2019transfer}. Therefore, the process of transferring knowledge from the source domain to the target domain is transformed to the data distribution matching. For instance, TCA \cite{pan2010domain} introduce the marginal maximum mean discrepancy (MMD) into the loss function, and then minimize the distance between the source and target features in regenerated Hilbert kernel space (RHKS). Existing works reveal that deep neural networks can disentangle more transferable representations than shallow models \cite{bengio2013representation,glorot2011domain}, but deep latent representations could only narrow but not eliminate the discrepancy of two domains \cite{yosinski2014transferable}. Inspired by adversarial learning \cite{lowd2005adversarial}, researchers mainly focus on adversarial domain adaptation methods, which align the source domain distribution and target domain distribution by adding adversarial objects into domain adaptation networks. When the adversarial object confuses the two domains, it is considered that the shared knowledge is adapted from the source to the target \cite{tzeng2015simultaneous}. 

Adversarial domain adaptation methods often utilize a domain classifier as the adversarial object. For example, the DANN \cite{ganin2016domain} is composed of a feature extractor, a label predictor and a domain classifier. The domain classifier tries its best to judge input features' source, but the feature extractor manages to deceive domain classifier with learned features. So DANN eliminates the domain discrepancy through the min-max game of the domain classifier and feature extractor. With the utilization of a gradient reversal layer (GRL) in the domain classifier, all the submodules are trained jointly to fulfill knowledge transfer. Unlike DANN, the ADDA \cite{tzeng2017adversarial} separately learns two feature extractors for the source domain and target domain. Specifically, the source feature extractor and label predictor are trained by minimizing the cross-entropy loss, while the target feature extractor and domain classifier are trained in adversarial manner when source feature extractor's parameters are fixed.

Different from existing works, this paper focuses on the detection of fake news in new events. We use adversarial adaptation networks to automatically extract meta knowledge for fact checking in new events, and we utilize a weighting mechanism to calculate the importance of each social post to alleviate negative transfer. 

\begin{figure}
\includegraphics[width=\textwidth]{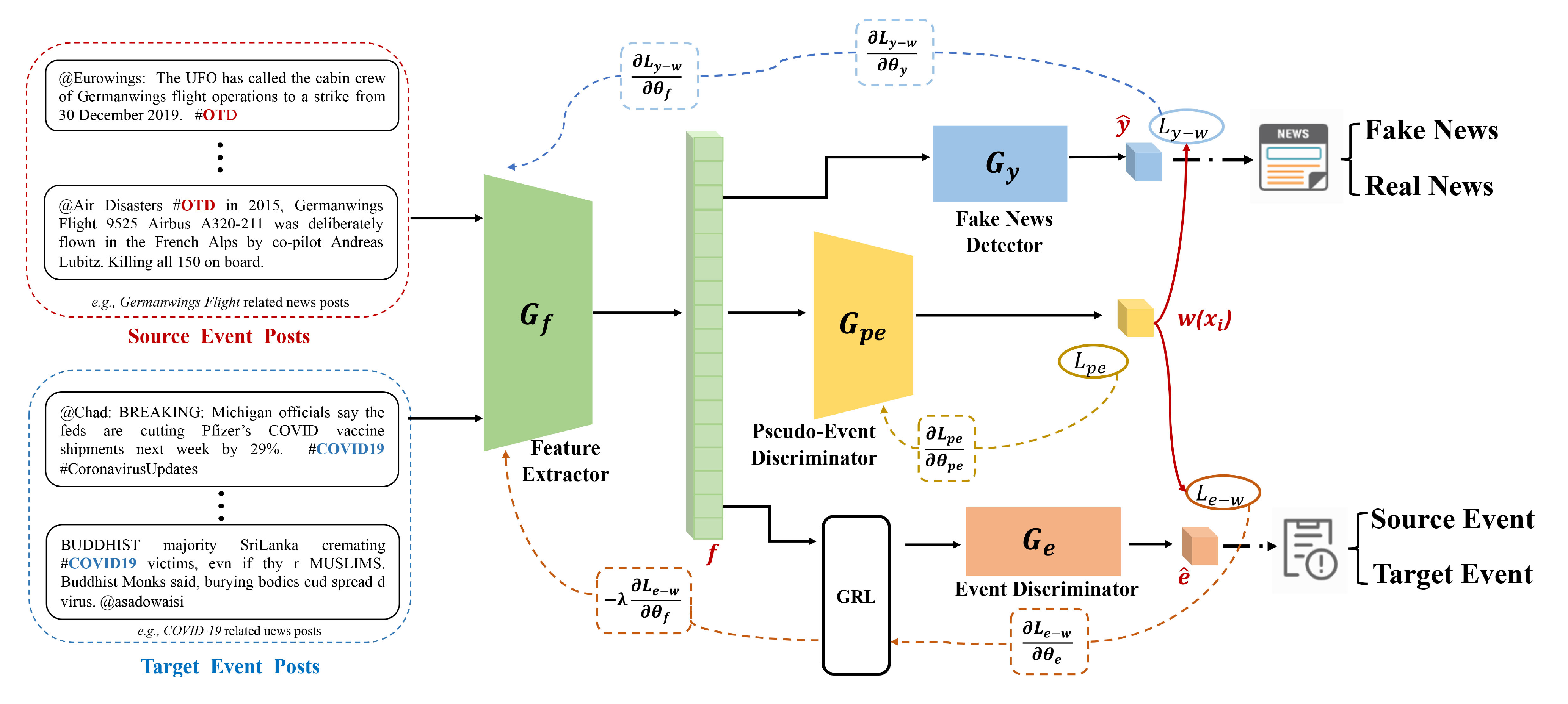}
\caption{The overall architecture of the \textit{MetaDetector}. It consists of four parts: (1) the feature extractor $G_{f}$ with parameters $\theta_{f}$, (2) the fake news detector $G_{y}$ with parameters $\theta_{y}$, (3) the pseudo-event discriminator $G_{pe}$ with parameters $\theta_{pe}$, and (4) the event discriminator $G_{e}$ with parameters $\theta_{e}$. Besides, the black solid lines represent the forward propagation, and the colored dashed lines represent the gradient back propagation.} \label{MetaDetector}
\end{figure}

\section{Methodology}
\label{Methodology}
In this section, we provide the problem definition of fake news, and present the framework in detail to show how to learn meta knowledge and utilize it to detect fake news in new events. 
\subsection{Problem Statement}
Given a source event dataset $\mathbb{D}_{s}=\left \{ (x_{i}^{s},y_{i}^{s}) \right \}_{i=1}^{n_{s}}$ with $n_{s}$ labeled posts (user published to describe the event or express viewpoints) and a target event dataset $\mathbb{D}_{t}=\left \{ x_{j}^{t} \right \}_{j=1}^{n_{t}}$ with $n_{t}$ unlabeled posts. The source and the target events are sampled from joint distributions $p_{s}\left ( \textbf{x}^{s},\textbf{y}^{s} \right )$ and $p_{t}\left ( \textbf{x}^{t},\textbf{y}^{t} \right )$ respectively (satisfying $p_{s}\neq p_{t}$). This paper aims to design a 
neural network $G:\textbf{x} \mapsto \textbf{y} $ to minimize the target cost $\delta _{t}\left ( G \right ) = \mathbb{D}_{\left ( \textbf{x}^{t},\textbf{y}^{t} \right )\sim p_{t}}\left [ G\left ( \textbf{x}^{t}\neq \textbf{y}^{t} \right ) \right ]$, which formally captures the transferable features of different events, reduces the event distribution discrepancy, and enhances the model generality to new events for identifying falsehood. We formalize the fake news detection task as follows:

\begin{definition}[Fake News Detection]
Under the supervised learning of source event data, fake news detection is defined as a binary classification task to predict whether a post $x_{j}^{t}\left ( j=1,2,\cdots ,n_{t} \right )$ of the target event is real or fake.
$$
G\left ( x_{j}^{t} \right )=
\begin{cases}
0, & \text{if $x_{j}^{t}$ is a fake post} \\
1, & \text{otherwise}
\end{cases}
$$
\end{definition}

\subsection{The MetaDetector Framework}
We propose the \textit{MetaDetector} for fake news detection in the context of historical knowledge transfer (as shown in Figure \ref{MetaDetector}), which mainly consists of the \textit{feature extractor}, the \textit{event discriminator}, the \textit{pseudo-event discriminator}, and the \textit{news detector}. Specifically, the workflow of our proposed model is as follows:
\begin{itemize}
    \item The \textit{feature extractor} embeds the source and target event data into a feature space.
    \item The \textit{pseudo-event discriminator} calculates the importance of source event posts to new events.
    \item The \textit{event discriminator} learns meta knowledge of different events by adversarial training. 
    \item The \textit{news detector} predicts the label of target event posts based on learned meta knowledge.
\end{itemize}

\subsubsection{Feature Extractor}
It maps news articles into dense vectors, extracts their semantic features, and passes them to subsequent modules. The feature extractor is implemented via Text-CNN  \cite{yoon2014convolutional}. For each word in both source and target posts, we utilize Word2Vec \cite{mikolov2013distributed} to train word embeddings, and then concatenate them as the initial representation of an entire post. Let $x_{i}\in \mathbb{R}^{d\times k}$ denotes the original representation of the \textit{i-th} input post, where $d$ is the word embedding dimension and $k$ is the max sentence length. Taking into account the efficiency and quality of feature extracting,\textit{ MetaDetector} uses Text-CNN to learn local features between words and phrases for describing linguistic characteristics of social posts.

The detailed process of content feature extraction is as follows. In the convolutional phase, each convolutional filter ($h$ in height and $d$ in width) takes embedding vectors of $h$ words as input and generates a corresponding feature $c_{i}$. Therefore, each current filter generates a feature vector $C_{i}=\left [ c_{1},c_{2},\cdots ,c_{k-h+1} \right ]$ ($C_{i}\in \mathbb{R}^{k-h+1}$) on post $x_{i}$. The max-pooling layer is able to select the maximum value in $C_{i}$ as a single feature extracted by the convolutional filter from the input post, i.e., $\hat{c_{i}}=max\left \{ C_{i} \right \}$. A variety of multi-granularity features can be learned by using several convolutional filters with different window sizes, so we set $w$ window sizes and adopt $n_{c}$ filters for each specific window size. The output of max-pooling layer is the concatenation of all calculated features, denoted as $
\hat{\textbf{c}}_{temp}\in \mathbb{R}^{w\cdot n_{c}}$. In order to integrate different features and fix the feature dimension, we further feed $\hat{\textbf{c}}_{temp}$ into an activation function. Finally, the output of feature extractor is:
\begin{equation}
\hat{\textbf{c}}=ReLU\left ( W_{fc}\cdot \hat{\textbf{c}}_{temp} + b_{fc} \right )
\end{equation}
where $W_{fc}$ and $b_{fc}$ are the weight matrix and bias of this fully connected layer, respectively. In summary, we abbreviate the feature extractor as $G_{f}\left ( \textbf{x} \right )$, i.e., $\hat{\textbf{c}}=G_{f}\left (\textbf{x} \right)$.

\subsubsection{Fake News Detector}
It judges whether the input post is real or fake news based on corresponding features $\hat{\textbf{c}}$ passed by the feature extractor, consisting of a single fully connected layer with softmax. We denote fake news detector as $G_{y}\left ( \cdot \right )$, and its output is:

\begin{equation}
\hat{\textbf{y}}=G_{y}\left(G_{f}\left(\textbf{x}\right)\right)
\end{equation}
where $\hat{\textbf{y}}$ is a tensor of size $n_{s} \times 2$ (indicating the probability that the post is a piece of fake news). Since posts in the source event are labeled, we can train $G_{y}$ with supervised classification cross-entropy loss $L_{y}$:  
\begin{equation}
L_{y}=\mathbb{E}_{\textbf{x},y \sim p_{s}\left(\textbf{x},y\right)}\left[y\log G_{y}\left( G_{f}\left(\textbf{x}\right)\right) + \left(1-y\right)\log\left(1-G_{y}\left(G_{f}\left(\textbf{x} \right)\right)\right)\right] \label{news_detection_loss}
\end{equation}

Current fake news detector optimized by minimizing loss $L_{y}$ can truly identify fake news in source event (i.e., $\min \limits_{G_{f},G_{y}} L_{y}$), while may be invalidate when facing the unseen events in the training stage. In other words, the cooperation between feature extractor and fake news detector lays great stress on event-specific features, ignoring transferable features that are instructive for the target event fake news detection. In order to adapt the fake news detector to target events samples, we need to acquire the meta knowledge underlying different events, correctly measure the dissimilarity between the source and target event, and combine learned meta knowledge and target event-specific features to detect fake news in unlabeled new events. 
\subsubsection{Event Discriminator}
The event discriminator is the function to measure the dissimilarity between the source and target event, which is the same as Wang et al. \cite{wang2018eann} and is composed of two fully connected layers. It takes features $\hat{\textbf{c}}$ as input, and outputs the probabilities that  the corresponding posts belong to the source and target event. The higher the probability is, the more likely the current feature approaches to event-specific features. In order to obtain the meta knowledge, it should be reduced from the mapped feature space. We denote the event discriminator as $G_{e}\left ( \cdot \right )$ and its final output $\hat{\textbf{e}}$ is:
\begin{equation}
\hat{\textbf{e}}=G_{e}\left(G_{f}\left(\textbf{x}\right)\right) 
\end{equation}
Assume that source event posts are positive samples (i.e., the event label is ``1'') and target event posts are negative samples (i.e., the event label is ``0''), this event discriminator loss $L_{e}$ can be described by:
\begin{equation}
 L_{e}=\mathbb{E}_{\textbf{x}\sim p_{s}\left(\textbf{x}\right)}\left [\log G_{e}\left(G_{f}\left( \textbf{x}\right)\right)\right] + \mathbb{E}_{\textbf{x}\sim p_{t}\left(\textbf{x}\right)}\left [\log \left ( 1-G_{e}\left(G_{f}\left( \textbf{x}\right)\right) \right )\right] \label{event_discrimination_loss}
\end{equation}

The event discriminator loss $L_{e}$ reflects the proximity of deep feature distributions of the source and target event. So the larger $L_{e}$ illustrates that event discriminator can not distinguish the source of input posts, which means that the combination of feature extractor and event discriminator can gradually reduce the discrepancy across events. Aiming at disentangling transferable features from the feature space as much as possible, the detection task plays a min-max game, which poses a challenge to model training by stochastic gradient descent (SGD): the feature extractor manages to confuse the event discriminator to increase $L_{e}$ scores, while the event discriminator confronts to clarify 
the source of news posts (source event or target event) to decrease $L_{e}$ scores (i.e., $\min \limits_{G_{f}}\max \limits_{G_{e}} L_{e}$).

Given features $\hat{\textbf{c}}=G_{f}\left (\textbf{x} \right)$ learned by feature extractor, the optimal event discriminator $G_{e}^{*}$ is obtained at:
\begin{equation}
    G_{e}^{*}\left(\hat{\textbf{c}}\right)=\frac{p_{s}\left ( \hat{\textbf{c}} \right )}{p_s{\left ( \hat{\textbf{c}} \right )} + p_{t}\left ( \hat{\textbf{c}} \right )} \label{optimal}
\end{equation}
According to Equation \ref{optimal}, if $G_{e}^{*}\rightarrow 1$, the current event discriminator can easily identify news posts from the source event. If $G_{e}^{*}$ is small enough, the $\hat{\textbf{c}}$ is most likely from the shared feature distribution of events. Similar to Zhang et al. \cite{zhang2018importance}, we give the following proof of Equation \ref{optimal}:
\begin{proof}
Given different event posts $\textbf{x}$, $G_{e}$ uses maximizing Equation \ref{event_discrimination_loss} as the training criterion:
\begin{equation}
\max_{G_{e}} L_{e}\left(G_{e}, G_{f}\right)=\int_{\textbf{x}} p_{s}(\textbf{x})\log G_{e}\left(G_{f}\left(\textbf{x}\right)\right)+p_{t}\left(\textbf{x}\right)\log \left ( 1-G_{e}\left ( G_{f}\left ( \textbf{x} \right ) \right ) \right ) d{\textbf{x}} \label{max_event_loss}
\end{equation}
Since feature distributions of the source and target event are both in the same feature space, we substitute original data distributions in Equation \ref{max_event_loss} with feature distributions, and calculate the partial differentiation with respective to $G_{e}$:
\begin{equation}
\begin{split}
    \frac{\partial L_{e}\left(G_{e},G_{f}\right)}{\partial G_{e}}&=\frac{\partial\int_{\hat{\textbf{c}}}p_{s}\left(\hat{\textbf{c}}\right)\log G_{e}\left(\hat{\textbf{c}}\right)+p_{t}\left(\hat{\textbf{c}}\right)\log\left(1-G_{e}\left(\hat{\textbf{c}}\right) \right)d\hat{\textbf{c}}}{\partial G_{e}} \\
    &= \int_{\hat{\textbf{c}}} p_{s}\left(\hat{\textbf{c}}\right)\cdot \frac{1}{G_{e}\left(\hat{\textbf{c}}\right)}-p_{t}\left( \hat{\textbf{c}}\right)\cdot \frac{1}{1-G_{e}\left(\hat{\textbf{c}}\right)}d\hat{\textbf{c}}
\end{split}
\end{equation}
According to the \textit{first-order optimality conditions}, we let $\partial L_{e}\left(G_{e},G_{f}\right) / \partial G_{e} = 0$, and the optimal event discriminator $G_{e}^{*}$ satisfies:
\begin{equation}
    p_{s}\left(\hat{\textbf{c}}\right)\cdot \frac{1}{G_{e}^{*}\left(\hat{\textbf{c}}\right)}-p_{t}\left(\hat{\textbf{c}} \right)\cdot \frac{1}{1-G_{e}^{*}\left(\hat{\textbf{c}}\right)}=0 \label{condition}
\end{equation}
Finally, the optimal event discriminator can be obtained by Equation \ref{condition}.
\end{proof}

\subsubsection{Pseudo-Event Discriminator}
It is responsible for measuring the importance of each source event post to the target event, just like a twin event discriminator, which is composed of two fully connected layers and denoted as $G_{pe}(\cdot)$. The input of the pseudo-event discriminator is features $\hat{\textbf{c}}$ fed by feature extractor, and the output is importance scores of source posts. Calculating the importance of source event posts is actually to align the source event and target event posts more accurately in the same feature space by re-weighting source event posts. To solve this problem, we apply the pseudo-event discrimination mechanism to measure needed weights. 

The mechanism is based on the following basic assumption: \textit{the source probability of each post calculated by event discriminator also reflects the degree of sharing between feature distributions of the source and target event}. Then meta knowledge can be disentangled based on some mathematical transformation of $\hat{\textbf{e}}$. This is why the mechanism is considered to be a pseudo-event discrimination mechanism, that is, the source posts' weights are defined as a function  similar to event discriminator. According to Equation \ref{optimal}, when $G_{e}^{*}$ approaches 1 ($p_{t}\left(\hat{\textbf{c}}\right)$ gradually decreasing), feature extractor pays more attention to source event-specific features, hence this type of posts should be given smaller weights. When $G_{e}^{*}$ is small enough ($p_{t}\left(\hat{\textbf{c}}\right)$ gradually increasing), feature extractor captures event-shared features required by fake news detector, and this kind of posts should be given larger weights. Based on the trade-off relationship between $G_{e}^{*}$ and weights $w^{*}$, we concisely define $w^{*}$ as follows:
\begin{equation}
    w^{*} = 1 - G_{e}^{*} \label{optimal_weight}
\end{equation}

Now that event discriminator can also measure the source post importance indirectly, the reason why we additionally introduce a pseudo-event discriminator is as follows: the introduced weight $w^{*}$ cannot theoretically reduce the feature distribution discrepancy between events in the same optimization. Since $w^{*}$ is also a function of $G_{e}^{*}$, the final optimal event discriminator is no longer being the ratio between the source feature distribution and the sum of the source and target feature distributions (as shown in Equation \ref{optimal}). 

Compared with $G_{e}$, the $G_{pe}$ does not undergo adversarial training, and its gradient does not need to be back-propagated to optimize $G_{f}$, considering that the gradient calculated in unweighted source posts and target posts cannot exactly reflect the corresponding event-shared feature distribution. Consequently, the $G_{pe}$ is used to calculate importance of source posts, and the $G_{e}$ is utilized to reduce the discrepancy in the feature distributions for preserving meta knowledge. According to Equation \ref{optimal_weight}, we denote output of $G_{pe}$ as $\hat{\textbf{w}}=G_{pe}\left(G_{f}\left(\textbf{x}\right)\right)$ ($\textbf{x}\in\left\{D_{s}\cup D_{t}\right\}$), and formalize the weights of source event posts ($\textbf{x}\in D_{s}$) as follows:
\begin{equation}
    w\left(\textbf{x}\right)= 1 - \hat{\textbf{w}} = 1 - G_{pe}\left(G_{f}\left(\textbf{x}\right)\right)
\end{equation}
Afterwards, the pseudo-event discriminator $G_{pe}$ can be trained with the following cross-entropy loss $L_{pe}$, and weights of source posts can be calculated by $\min \limits_{G_{pe}} L_{pe}$: 
\begin{equation}
 L_{pe}=\mathbb{E}_{\textbf{x}\sim p_{s}\left(\textbf{x}\right)}\left [\log G_{pe}\left(G_{f}\left( \textbf{x}\right)\right)\right] + \mathbb{E}_{\textbf{x}\sim p_{t}\left(\textbf{x}\right)}\left [\log \left ( 1-G_{pe}\left(G_{f}\left( \textbf{x}\right)\right) \right )\right] \label{weight_evaluation_loss}
\end{equation}

\subsection{The Loss Function of MetaDetector}
We now introduce the loss function and working framework of our proposed \textit{MetaDetector}. While optimizing a fake news detecting objective, it learns cross-event meta knowledge for verifying unlabeled target event posts, which is performed by concurrently optimizing the supervised fake news detection loss $L_{y}$, supervised event discrimination loss $L_{e}$, and supervised weight evaluation loss $L_{pe}$. 

On the basis of event-level adversarial adaptation, \textit{MetaDetector} utilizes the pseudo-event discrimination-based weighting mechanism to match the key feature distributions of events, which reduces the impact of irrelevant or anomalous source posts on the unlabeled target posts, and uses the learned meta knowledge to detect fake news in new events, which facilities post-level adversarial adaptation. In practical scenarios, sometimes the feature distribution of the source and target event are very comparable. For example, some new fake news just replace names and locations mentioned in historical fake news. Limited by the size of source event data for training \textit{MetaDetector}, the weights applied in such cases explicitly weaken the representations of transferable features to a certain extent (since each $w\left(x_{i}\right)$ is a scalar between 0 and 1), which may lead to the decline of detecting accuracy. Consequently, \textit{MetaDetector} automatically determines the values of source posts' weights according to the distribution discrepancy $d_{k}$ between the source event and target event, i.e.,    
\begin{equation}
w\left ( \textbf{x} \right )=
\begin{cases}
1-G_{pe}\left(G_{f}\left(\textbf{x}\right)\right), & \text{if $d_{k}$ $\geq$ $d^{*}$} \\
\left[1,1,\cdots,1 \right]^{T}_{1 \times n_{s}}, & \text{otherwise}
\end{cases}    
\end{equation}
where $d^{*}$ is a hyper-parameter of event distribution shift threshold and $d_{k}$ is calculated by the maximum mean discrepancy (MMD) \cite{long2015learning}. The squared form of $d_{k}$ is defined as:
\begin{equation}
    d_{k}^{2}\left ( p_{s}, p_{t} \right )=\left \| E_{\textbf{x}\sim p_{s}\left (\textbf{x}\right)}\left [\phi \left(\textbf{x}\right)\right] - E_{\textbf{x}\sim p_{t}\left( \textbf{x}\right)}\left [ \phi \left(\textbf{x}\right) \right ] \right \|_{\mathbb{H}}^{2}
\end{equation}
where $\phi \left(\cdot\right)$ is a feature mapping function and the two expectations represent the center of the source and target feature distribution respectively.

For transferring meta knowledge between events to detect fake news in new events, we re-formulate the fake news detection loss and event discrimination loss by adding weights to Equation \ref{news_detection_loss} and \ref{event_discrimination_loss} respectively. Specifically, the weighted fake news detection loss $L_{y-w}$ and weighted event discrimination loss $L_{e-w}$ are as follows:
\begin{equation}
   L_{y-w}=\mathbb{E}_{\textbf{x},y \sim p_{s}\left(\textbf{x},y\right)} w\left (\textbf{x}\right )\left[y\log G_{y}\left( G_{f}\left(\textbf{x}\right)\right) + 
   \left(1-y\right)\log\left(1-G_{y}\left(G_{f}\left(\textbf{x} \right)\right)\right)\right] \label{weighted_news_detection_loss} 
\end{equation}
\begin{equation}
   L_{e-w}=\mathbb{E}_{\textbf{x}\sim p_{s}\left(\textbf{x}\right)}\left [w\left(\textbf{x}\right)\log G_{e}\left(G_{f}\left( \textbf{x}\right)\right) \right] + 
  \mathbb{E}_{\textbf{x}\sim p_{t}\left(\textbf{x}\right)}\left [\log \left ( 1-G_{e}\left(G_{f}\left( \textbf{x}\right)\right) \right )\right] \label{weighted_event_discrimination_loss}
\end{equation}

In conclusion, \textit{MetaDetector} optimizes the fake news detector by minimizing the weighted fake news detection loss $L_{y-w}$, learns the event discriminator by minimizing the weighted event discrimination loss $L_{e-w}$, and cultivates the pseudo-event discriminator by minimizing the weight evaluation loss $L_{pe}$. Besides, the feature extractor is optimized to minimize $L_{y-w}$ but maximize $L_{e-w}$ at the same time. The final loss of \textit{MetaDetector} is the linear combination of all three losses:
\begin{equation}
    L_{final} = L_{y-w}+\mu L_{w}-\lambda L_{e-w} 
\end{equation}
where $\mu$ is a scalar that adjusts the impact of weight evaluation loss on the final loss function, and $\lambda$ is a hyper-parameter that adapts to the trade-off between the fake news detection and event discrimination. To perform the min-max game between feature extractor and event discriminator, we add the GRL \cite{ganin2015unsupervised} in the event discriminator before fully connected layers, which has no other parameters except the hyper-parameter $\lambda$. Specifically, during forward propagation, GRL is equivalent to an identity transformation function, which feeds the features extracted from $G_{f}$ into $G_{e}$. During back propagation, GRL receives the gradient from its subsequent layer, multiplies it by $-\lambda$, and passes it to $G_{f}$ for learning parameters of $G_{f}$ and $G_{e}$ simultaneously. 

\section{Experiments}
\label{Experiments}
In this section, we conduct extensive experiments on two large-scale datasets, and compare the performance of \textit{MetaDetector} against the baselines. In addition, we perform case study to evaluate the effectiveness of \textit{MetaDetector}.

\subsection{Datasets}
To evaluate the performance of the proposed solution, we collected two large-scale fake news dataset from Sina Weibo and Twitter. We perform several fake news detection tasks by transferring shared features from $A \rightarrow B$, where $A$ corresponds to the source events and $B$ represents the target events.

\subsubsection{Weibo} For collecting enough news data, we have integrated three existing datasets to a single weibo dataset, including:
\begin{itemize}
    \item The first part is extracted all the textual modal content from the data released by Jin et al. \cite{jin2017multimodal} that contains 4,779 real news and 4,749 fake news, crawled from May, 2012 to January, 2016.  
    \item The second part is published by Ma et al. \cite{ma2016detecting} that includes 2,313 fake news and 2,351 real news. This dataset also collects original news posts, as well as their comments. Similar to the first part, we apply the original text of this data.
    \item The third part is a COVID-19 fake news dataset published by Yang et al. \cite{yang2020checked}, namely CHECKED, which consists of 344 fake news and 1,776 real news about COVID-19 collected from December, 2019 to August, 2020.
\end{itemize}
Since the first two parts of our dataset do not distinguish the association between news, we perform hierarchical clustering on them to characterize upcoming events after removing posts less than 10 in length, similar to Wang et al. \cite{wang2018eann}. Then we select 3 generalized events from clustering results (abbreviated as $Evt_{1}$, $Evt_{2}$, and $Evt_{3}$) and the COVID-19 news data. Afterwards, we take each event as the source or target event to define six fake news detection tasks: $Evt_{1}\rightarrow Evt_{2}$, $Evt_{2}\rightarrow Evt_{1}$, $Evt_{1}\rightarrow Evt_{3}$, $Evt_{3}\rightarrow Evt_{1}$, $Evt_{2}\rightarrow Evt_{3}$, and $Evt_{3}\rightarrow Evt_{2}$.

\subsubsection{Twitter} We select the tweets of three breaking news from PHEME \cite{zubiaga2016learning} to formalize fake news detection tasks. The three representative events are \textit{Charlie Hebdo}\footnote{https://en.wikipedia.org/wiki/Charlie\_Hebdo\_shooting} (denoted as $CH$), \textit{Ferguson}\footnote{https://en.wikipedia.org/wiki/Shooting\_of\_Michael\_Brown} (denoted as $Fe$), and \textit{Sydney Siege}\footnote{https://en.wikipedia.org/wiki/Lindt\_Cafe\_siege} (denoted as $SS$). The six fake news detection tasks are formalized as follows: $CH\rightarrow Fe$, $Fe\rightarrow CH$, $CH\rightarrow SS$, $SS\rightarrow CH$, $Fe\rightarrow SS$, and $SS\rightarrow Fe$. 

The statistics of the two datasets are shown in Table \ref{data_statistics}. For each detection task of Weibo, we randomly select 2,500 labeled source posts and 2,500 unlabeled target posts for training, 500 target posts for validation, and 900 target posts for testing. For the Twitter data, we choose 800 labeled source tweets and 800 unlabeled target tweets for training, 150 target tweets for validation, and target tweets for testing ranged from 200 to 300.

\begin{table}
\caption{Statistics of Datasets} \label{data_statistics}
\scriptsize
\centering
\setlength{\tabcolsep}{3.0mm}{
\begin{tabular}{ccccc}
\toprule
\multirow{2}{*}{\textbf{Data Source}}    &  \multirow{2}{*}{\textbf{Event}}    &  \multicolumn{3}{c}{\textbf{Number of News Posts}}    \\
\cmidrule(r){3-5}               &                & \textbf{Real}   & \textbf{Fake}   & \textbf{Total}  \\
\midrule
\multirow{4}{*}{Sina Weibo}     & $Evt_{1}$      & 2,704           & 1,400           & 4,104        \\
                                & $Evt_{2}$      & 962             & 2,938           & 3,900        \\
                                & $Evt_{3}$      & 2,459           & 1,478           & 3,937        \\
                                & COVID-19       & 1,776           & 344             & 2,120         \\
\midrule
\multirow{3}{*}{Twitter}        & Charlie Hebdo  & 1,621           & 458             & 2,079         \\
                                & Ferguson       & 859             & 284             & 1,143         \\
                                & Sydney Siege   & 699             & 522             & 1,221         \\
\bottomrule
\end{tabular}}
\end{table}

\subsection{Implementation Details}
\subsubsection{Baselines} 
We compare our proposed model with the following representative methods:
\begin{itemize}
    \item DNN: It uses the typical nonlinear fitting ability of multiple fully connected layers to learn the textual features of news. In this experiment, the DNN model adopts two fully connected layers, and the size of the first layer and the second layer are 64 and 32, respectively.
    \item Text-CNN \cite{yoon2014convolutional}: It uses convolutional neural networks to model content in news, which obtains latent linguistic features of different granularities by diverse filters. The initial parameter setting is the same as the $G_{f}$ (see details in section \ref{experimental_settings})
    \item GRU-2 \cite{ma2016detecting}: It takes the news content as a time sequential data, and then applies two GRU hidden layers to obtain high-level feature representations of the news.
\end{itemize}

In addition, to study the performance of the \textit{MetaDetector} on the upcoming news, we also take following advanced domain adaptation methods (adjusted to fake news detection tasks) as the baselines.
\begin{itemize}
    \item EANN \cite{wang2018eann}: It jointly trains a multi-modal feature extractor, an event discriminator, and a fake news detector for multi-modal fake news detection. Since this article focuses on the text modal, we remove the feature extraction part of the visual modal and denote it as \textit{EANN-text}.
    \item ADDA \cite{tzeng2017adversarial}: It trains a feature extractor for the source domain and target domain respectively, and combines discriminative modeling, untied weight sharing, and a GAN loss for characterizing representations of the target domain in the shared feature space. In order to make a fair comparison with our pseudo-event discrimination-based weighting mechanism, we utilize the ADDA model based on the GRL layer (abbreviated as \textit{ADDA-grl}), where the min-max game is not iteratively trained by a GAN loss but is implemented through the GRL layer. Besides, both the source and target encoders of ADDA-grl are Text-CNNs.
\end{itemize}

Note that unlabeled target event instances are not used when training non-transfer fake news detection methods, i.e., the DNN, Text-CNN, and GRU-2.  

\begin{table}
\caption{The MMD Distance Between Different Events} \label{MMD_distance_table}
\scriptsize
\centering
\setlength{\tabcolsep}{2.0mm}{
\begin{tabular}{cccc}
\toprule
\textbf{Data Source}    &  \textbf{Source}    &  \textbf{Target} &   \textbf{MMD Distance Between Events} \\
\midrule
\multirow{3}{*}{Twitter}    & $CH$      & $Fe$      & 0.6378        \\
                            & $CH$      & $SS$      & 0.1734        \\
                            & $Fe$      & $SS$      & 0.3728        \\
\midrule
\multirow{6}{*}{Sina Weibo}     & $Evt_{1}$           & $Evt_{2}$   & 1.3567 \\
                                & $Evt_{1}$           & $Evt_{3}$   & 0.3386 \\
                                & $Evt_{2}$           & $Evt_{3}$   & 1.3315 \\
                                & \textit{COVID-19}   & $Evt_{1}$   & 0.4905 \\
                                & \textit{COVID-19}   & $Evt_{2}$   & 1.1324 \\
                                & \textit{COVID-19}   & $Evt_{3}$   & 0.9161 \\
\bottomrule
\end{tabular}}
\end{table}

\subsubsection{Experimental settings}
\label{experimental_settings}
For training the \textit{MetaDetector}, we set $\lambda = \mu = 1$ for the final loss function $L_{final}$, calculate the MMD distance between events using 7 Gaussian kernels, and set the distribution shift threshold $d^{*}$ to 0.8 according to calculated distances in Table \ref{MMD_distance_table}. In the feature extractor, we set the dimension of word embeddings $d=32$, and the maximum sentence length $k$ depends on posts' lengths in Sina Weibo and Twitter data respectively. In addition, the number of convolutional filters $n_{c}$ is 20, and its window size $w$ ranges from 1 to 4. The hidden size of the fully connected in this feature extractor is 32. The event discriminator and the pseudo-event discriminator are the same architecture, consisting of two fully connected layers. In the fake news detector, the hidden size of the single fully connected layer is also 32. For all the baselines and the proposed model, the training batch size and number of epochs are both 100, learning rate is 0.01, and the dropout rate is 0.2.

\subsection{Fake News Detection Results}
\begin{figure}
\includegraphics[width=\textwidth]{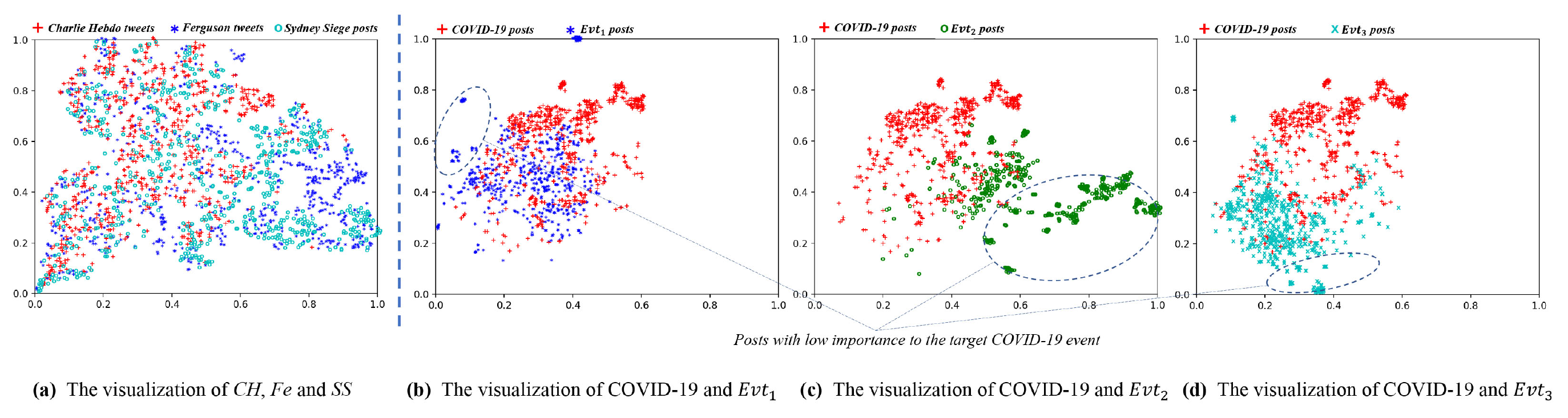}
\caption{The t-SNE visualization of data distributions in our two experimental datasets. In sub-figure(a), the red nodes are \textit{CH} tweets, blue nodes are \textit{Fe} tweets, and cyan nodes are \textit{SS} tweets. The red nodes in sub-figure(b) to (d) are all \textit{COVID-19} related posts, and the blue, green, cyan nodes represent $Evt_{1}$ posts, $Evt_{2}$ posts and $Evt_{3}$ posts, respectively.} 
\label{data_tsne}
\end{figure}

We first use t-SNE to visualize the distribution shifts among the three  breaking news (\textit{Charlie Hebdo Shooting}, \textit{Ferguson Unrest}, and \textit{Sydney Siege}) in Twitter. As shown in Figure \ref{data_tsne}(a), the distribution shifts among the three events are subtle. This could be resulted from the fact that the three events are all related to shooting or armed robbery crimes. Apart from the event properties including the locations and scenarios, the text descriptions of events are semantic comparable, which are suitable for detecting fake news by transferring event meta knowledge. Figure \ref{data_tsne}(b)-(d) present the data distribution shifts between one specific event and the COVID-19 event. Each sub-figure illustrates that $Evt$ contains information that is abnormal or not shared with COVID-19 (samples circled by dotted lines), which should be given a smaller weight, and it also contains important transferable information for identifying posts in new events. Subsequently, we compare the performance of our proposed \textit{MetaDetector} and baseline models on designed fake news detection tasks.

Specifically, we compare the accuracy of \textit{MetaDetector} and baselines on a given set of detection tasks on Twitter (posts in English) and Sina Weibo (posts in Chinese), and design 6 knowledge transfer-based fake news detection tasks on each dataset. Table \ref{weibo_results} and Table \ref{twitter_results} respectively indicate the fake news detection results of different models. In addition, we take $Evt_{1}$, $Evt_{2}$, and $Evt_{3}$ as source events and the COVID-19 related posts as the target event to evaluate the detection performance of our \textit{MetaDetector}, mainly characterized by the \textit{accuracy} (abbreviated as $Acc.$), \textit{precision} (denoted as $P$), \textit{recall} (denoted as $R$), and \textit{F1 score}, as shown in Table \ref{COVID_results}. We can observe that our proposed \textit{MetaDetector} shows superior performance on most tasks, and it has competitive and even superior performance against the state-of-the-art domain adaptation methods.

\begin{table}
\caption{Accuracy (\%) on the Weibo dataset for fake news detection} 
\label{weibo_results}
\scriptsize
\centering
\renewcommand\arraystretch{1.5}{
\begin{tabular}{cccccccc}
\toprule
\multirow{2}{*}{\textbf{Method}}  & \multicolumn{6}{c}{\textbf{Fake News Detection Tasks in Weibo Dataset}}  & \multirow{2}{*}{\textbf{Avg.}}  \\
\cmidrule(r){2-7}    & $Evt_{1} \rightarrow Evt_{2}$    & $Evt_{1} \rightarrow Evt_{3}$   & $Evt_{2} \rightarrow Evt_{1}$   & $Evt_{2} \rightarrow Evt_{3}$  & $Evt_{3} \rightarrow Evt_{1}$ & $Evt_{3} \rightarrow Evt_{2}$  &  \\
\midrule
DNN       & 57.89 & 62.77 & 58.25 & 59.22 & 59.72 & 64.35     & 60.37  \\
GRU-2     & 68.78 & 64.29 & 65.99 & 67.80 & 60.83 & 66.67     & 65.72  \\
Text-CNN  & 76.78 & 63.69 & 69.67 & 71.38 & 65.56 & 78.67     & 70.96  \\
\cdashline{1-8}
EANN-text & 81.22 & 64.22 & 65.22 & 71.00 & 71.56 & 76.33     & 71.59  \\
ADDA-grl  & 81.78 & \textbf{67.68} & 70.69 & 71.63 & \textbf{73.17} & 78.04 & 73.83 \\
\cdashline{1-8}
MetaDetector & \textbf{82.67} & 65.56 & \textbf{71.38} & \textbf{75.56} & 72.63 & \textbf{81.00}  & \textbf{74.80}  \\
\bottomrule
\end{tabular}}
\end{table}

\begin{table}
\caption{Accuracy (\%) on the Twitter dataset for fake news detection} \label{twitter_results}
\scriptsize
\centering
\renewcommand\arraystretch{1.5}{
\begin{tabular}{cccccccc}
\toprule
\multirow{2}{*}{\textbf{Method}}  & \multicolumn{6}{c}{\textbf{Fake News Detection Tasks in Twitter Dataset}}  & \multirow{2}{*}{\textbf{Avg.}}  \\
\cmidrule(r){2-7}    & $CH \rightarrow Fe$    & $CH \rightarrow SS$   & $Fe \rightarrow CH$   & $Fe \rightarrow SS$  & $SS \rightarrow CH$ & $SS \rightarrow Fe$  & \\
\midrule
DNN        &  60.57     & 68.85      & 61.88      & 64.98      & 58.87      & 57.07      & 61.87  \\
GRU-2      &  63.82    & 72.47      & 64.22      & 63.35      & 61.17      & 59.02       & 64.34  \\
Text-CNN   &  64.65     & 75.86      & 65.97      & 66.44      & 65.58       & 62.23        & 65.79  \\
\cdashline{1-8}
EANN-text          &  66.67     & 75.52      & 68.18      & 71.15      & 65.29      & 67.68        & 69.08  \\
ADDA-grl           &  69.88     & \textbf{79.64}      & 69.48       & 70.08      & \textbf{66.88}       & 69.21        & 70.86  \\
\cdashline{1-8}
MetaDetector       &  \textbf{72.04}    & 78.85       & \textbf{72.69}  & \textbf{71.63}       & 66.52      &  \textbf{71.38}       & \textbf{72.19}  \\
\bottomrule
\end{tabular}}
\end{table}

\subsubsection{Effectiveness of meta knowledge transfer}
\label{effective_knowledge_transfer}
In the experimental results of two datasets, the fake news detection accuracy of knowledge transfer-based methods (i.e., \textit{EANN-text}, \textit{ADDA-grl}, and our proposed \textit{MetaDetector}) is generally higher than that of traditional deep learning models (i.e., \textit{DNN}, \textit{GRU-2}, and \textit{Text-CNN}). Compared with non-knowledge transfer methods, the average detection accuracy of \textit{MetaDetector} is improved by nearly 3\%-14\% in Weibo data tasks and 7\%-10\% in Twitter data tasks. In fact, these non-knowledge transfer methods do not fully consider the distribution discrepancy between the source event and the target event, and they are more sensitive to labeled source event posts, which leads to a special attention to the source event-specific features during the model training. Consequently, the ability to recognize fake news in new events declines dramatically. For example, for the task $CH \rightarrow Fe$, the MMD distance between the source $CH$ and target $Fe$ is 0.6378 (higher than the distances between other events in Twitter dataset), and  the detection accuracy of DNN and GRU-2 are approximately 10\% lower than that of \textit{MetaDetector}. In addition, the detection accuracy of Text-CNN is typically higher than that of GRU-2 and DNN, and is equivalent to the knowledge transfer methods especially in the task $Evt_{2} \rightarrow Evt_{3}$ and $Evt_{3} \rightarrow Evt_{2}$, which may be explained by the latent representation of fake news in this paper. \textit{MetaDetector} mainly explores the event-shared latent semantic information underlying the news contents of different events, so it does not combine comments or user profiles of disseminators to characterize each piece of news in new events (since this type of contextual information is mostly event-specific). The GRU-2 models textual information as time series data to capture discriminative features of the questionable real and fake news, which pays more attention to global semantic features and ignores local linguistic features. Therefore, the lack of feature representation for social reactions modeling has a certain negative impact on its detection accuracy. The Text-CNN relies on the multi-size convolutional filters to extract the rich feature representations of news posts, resulting in a 6.71\% higher accuracy rate than DNN in task $CH \rightarrow SS$, and 14.32\% higher in task $Evt_{1} \rightarrow Evt_{2}$.

\begin{table}
\caption{Fake News Detection Results from Historical Events to COVID-19} \label{COVID_results}
\scriptsize
\centering
\renewcommand\arraystretch{1.5}{
\begin{tabular}{c|c|c|c|c|c|c|c|c|c|c|c|c|c}
\hline
\multirow{2}{*}{\textbf{Method}}       & \multirow{2}{*}{\textbf{Class}} & \multicolumn{4}{c|}{\textbf{$Evt_{1}$$\rightarrow$\textit{COVID-19}}}                                                             & \multicolumn{4}{c|}{\textbf{$Evt_{2}$$\rightarrow$\textit{COVID-19}}}                                                             & \multicolumn{4}{c}{\textbf{$Evt_{3}$$\rightarrow$\textit{COVID-19}}}                                                             \\ \cline{3-14} 
                              &                        & Acc.                        & P      & R         & F1             & Acc.                        & P      & R         & F1             & Acc.                        & P      & R         & F1             \\ \hline \hline
\multirow{2}{*}{DNN}          & real                   & \multirow{2}{*}{0.647}          & 0.767          & 0.544          & 0.637          & \multirow{2}{*}{0.562}          & 0.601          & 0.700          & 0.647          & \multirow{2}{*}{0.617}          & 0.590          & 0.888          & 0.709          \\ \cline{2-2} \cline{4-6} \cline{8-10} \cline{12-14} 
                              & fake                   &                                 & 0.565          & 0.781          & 0.656          &                                 & 0.485          & 0.378          & 0.425          &                                 & 0.719          & 0.315          & 0.438          \\ \hline
\multirow{2}{*}{GRU-2}        & real                   & \multirow{2}{*}{0.712}          & 0.762          & 0.798          & 0.779          & \multirow{2}{*}{0.669}          & 0.674          & 0.732          & 0.702          & \multirow{2}{*}{0.711}          & 0.777          & 0.725          & 0.750          \\ \cline{2-2} \cline{4-6} \cline{8-10} \cline{12-14} 
                              & fake                   &                                 & 0.603          & 0.553          & 0.577          &                                 & 0.662          & \textbf{0.697} & \textbf{0.628} &                                 & 0.626          & 0.689          & 0.656          \\ \hline
\multirow{2}{*}{Text-CNN}     & real                   & \multirow{2}{*}{0.710}          & 0.777          & 0.725          & 0.750          & \multirow{2}{*}{0.677}          & 0.634          & 0.937          & 0.756          & \multirow{2}{*}{0.707}          & 0.771          & 0.741          & 0.756          \\ \cline{2-2} \cline{4-6} \cline{8-10} \cline{12-14} 
                              & fake                   &                                 & 0.626          & 0.689          & 0.656          &                                 & \textbf{0.839} & 0.377          & 0.520          &                                 & 0.565          & 0.606          & 0.585          \\ \hline \hline
\multirow{2}{*}{EANN-text}    & real                   & \multirow{2}{*}{0.820}          & 0.858          & 0.938          & 0.896          & \multirow{2}{*}{0.822}          & 0.845          & 0.962          & 0.899          & \multirow{2}{*}{0.818}          & 0.692          & 0.517          & 0.592          \\ \cline{2-2} \cline{4-6} \cline{8-10} \cline{12-14} 
                              & fake                   &                                 & 0.456          & 0.252          & 0.325          &                                 & 0.441          & 0.146          & 0.219          &                                 & \textbf{0.848} & 0.921          & 0.883          \\ \hline
\multirow{2}{*}{ADDA-grl}     & real                   & \multirow{2}{*}{\textbf{0.849}} & \textbf{0.897} & 0.827          & 0.861          & \multirow{2}{*}{0.835}          & 0.838          & \textbf{0.992} & 0.909          & \multirow{2}{*}{0.827}          & 0.737          & 0.500          & 0.596          \\ \cline{2-2} \cline{4-6} \cline{8-10} \cline{12-14} 
                              & fake                   &                                 & \textbf{0.796} & \textbf{0.876} & \textbf{0.834} &                                 & 0.667          & 0.078          & 0.139          &                                 & 0.845          & \textbf{0.939} & \textbf{0.889} \\ \hline
\multirow{2}{*}{MetaDetector} & real                   & \multirow{2}{*}{0.840}          & 0.846          & \textbf{0.986} & \textbf{0.911} & \multirow{2}{*}{\textbf{0.847}} & \textbf{0.865} & 0.966          & \textbf{0.913} & \multirow{2}{*}{\textbf{0.869}} & \textbf{0.863} & \textbf{0.889} & \textbf{0.875} \\ \cline{2-2} \cline{4-6} \cline{8-10} \cline{12-14} 
                              & fake                   &                                 & 0.667          & 0.136          & 0.226          &                                 & 0.622          & 0.272          & 0.378          &                                 & 0.844          & 0.811          & 0.827          \\ \hline
\end{tabular}}
\end{table}

\subsubsection{Effectiveness of the pseudo-event discrimination-based weighting mechanism}
\label{effective_weighting_mechanism}
Compared with other transfer learning methods, the average fake news detection accuracy of our proposed \textit{MetaDetector} on setting tasks of the two datasets is higher than that of baselines (74.8\% in Weibo data and 72.19\% in Twitter  data). In this experiment, we mainly compare our proposed model with \textit{EANN-text} and \textit{ADDA-grl}. As mentioned in Section \ref{Introduction}, EANN-text is an adversarial nets-based fake news detection method, which utilizes a domain classifier to guide its feature extractor to learn event-shared features. In fact, the adversarial training stage only reduces the marginal distribution discrepancy between events, and does not thoroughly align the conditional distribution of the source and target events. In other words, there is misleading information underlying learned event-shared features, and it matches the target posts to some specific source posts, making it perform weaker than Text-CNN on task $Evt_{2} \rightarrow Evt_{1}$ and $Evt_{3} \rightarrow Evt_{2}$, i.e., inducing the negative transfer. EANN-text can be regarded as a simplified version of our proposed model without the weighting mechanism. Therefore, we use the same network parameters as EANN-text to train \textit{MetaDetector}. The results in Table \ref{weibo_results} and Table \ref{twitter_results} indicate that \textit{MetaDetector} outperforms \textit{EANN-text} on our designed detection tasks. The average accuracy of \textit{MetaDetector} is 3.21\% higher than that of EANN-text in Weibo tasks, especially 6.16\% and 4.67\% in task $Evt_{2} \rightarrow Evt_{1}$ and $Evt_{3} \rightarrow Evt_{2}$. As shown in Table \ref{MMD_distance_table}, the MMD distances between $Evt_{1}$ and $Evt_{2}$, $Evt_{2}$ and $Evt_{3}$ are greater than the threshold $d^{*}$ (set by empirical analysis of events' distributions). In these two tasks, when \textit{MetaDetector} aligns the source event distribution with the target event distribution, it re-weights source posts through the importance scores learned by the pseudo-event discriminator, which reduces the negative impact of instances, similar to the circled nodes in Figure \ref{data_tsne}(c), on the target event fake news detection. Therefore, compared with the EANN-text, \textit{MetaDetector} does not show obvious negative transfer in 12 detection tasks. This demonstrates that the effectiveness of the pseudo-event discrimination-based weighting mechanism, which cultivates our proposed model to comprehensively evaluate the importance of source event posts, recognizes abnormal or unrelated historical instances, and reduces the discrepancy in feature distributions between the source event and an upcoming event.  

Although the accuracy of \textit{MetaDetector} is slightly lower than that of ADDA-grl in specific tasks (i.e., $CH \rightarrow SS$, and $Evt_{1} \rightarrow Evt_{3}$), the overall detection performance is comparable to ADDA-grl. In additional to no weighting mechanism, the ADDA-grl utilizes a non-shared feature extraction operation and uses a source encoder and a target encoder to extract the source event and target event features respectively. When the discrepancy of feature distributions is narrow, the non-shared feature extractor allows ADDA-grl to preserve the part of the transferable features that is closer to the target event, which explains the highest detection accuracy (2.12\% higher than our method) in task $Evt_{1} \rightarrow Evt_{3}$ (MMD distance is 0.3386). However, ADDA-grl matches the source event and target event distributions in two separate feature spaces. When the distribution discrepancy between the events is noticeable (the MMD distance is large enough), the accuracy of detecting fake news in target events is lower than that of \textit{MetaDetector}, which performs event meta knowledge transfer by re-weighting the source event distribution in the same feature space.

In order to illustrate the importance of event meta knowledge for detecting fake news in new events, we further utilize the \textit{MetaDetector} trained on historical events (i.e., $Evt_{1}$, $Evt_{2}$, and $Evt_{3}$) to identify fake news in the COVID-19 event \cite{yang2020checked}. Considering the small amount of high-quality COVID-19 posts, we sample 1,500 labeled source posts and 1,000 unlabeled epidemic posts for training, 400 and 600 epidemic posts for validating and testing respectively. Afterwards, we set the following fake news detection tasks: $
Evt_{1}\rightarrow$ \textit{COVID-19}, $Evt_{2}\rightarrow$ \textit{COVID-19}, and $Evt_{3}\rightarrow$ \textit{COVID-19}. We compare the detection accuracy, precision, recall and F1 scores of each fake news detection model under the three detection tasks, as shown in Table \ref{COVID_results}. It demonstrates that \textit{MetaDetector} has achieved the highest accuracy rate on the $Evt_{2}\rightarrow$ \textit{COVID-19} and $Evt_{3}\rightarrow$ \textit{COVID-19} tasks, which is equivalent to the highest accuracy rate on task $Evt_{1}\rightarrow$ \textit{COVID-19}. A noteworthy experimental result is that the recall, precision, and F1 scores of \textit{MetaDetector} on real news are all higher than other baselines, but it performs poorly on these metrics of fake news. For example, in task $Evt_{1}\rightarrow$ \textit{COVID-19}, the F1 scores of ADDA-grl on fake news is 0.834, while our model is only 0.226. Since the ratio of real and fake news in the COVID-19 data is close to 5:1, \textit{MetaDetector} may pay more attention to real news of the target event in the same feature space, resulting in a decline in the recall of fake news. Note that the MMD distance between $Evt_{1}$ and the \textit{COVID-19} (0.4095, details in Table \ref{MMD_distance_table}) is much smaller than our setting threshold, and their distribution discrepancy is much narrow. At this case, it is more conductive to the ADDA-grl that utilizes the non-shared feature extraction operation. For the other two tasks, as the distribution discrepancy increases, the MMD distance increasing from 0.9161 to 1.1324, we can observe that \textit{MetaDetector} maintains a high accuracy and F1 score in real news, while the precision, recall, and F1 scores of fake news are all equivalent to ADDA-grl. 

\begin{figure}
\includegraphics[width=\textwidth]{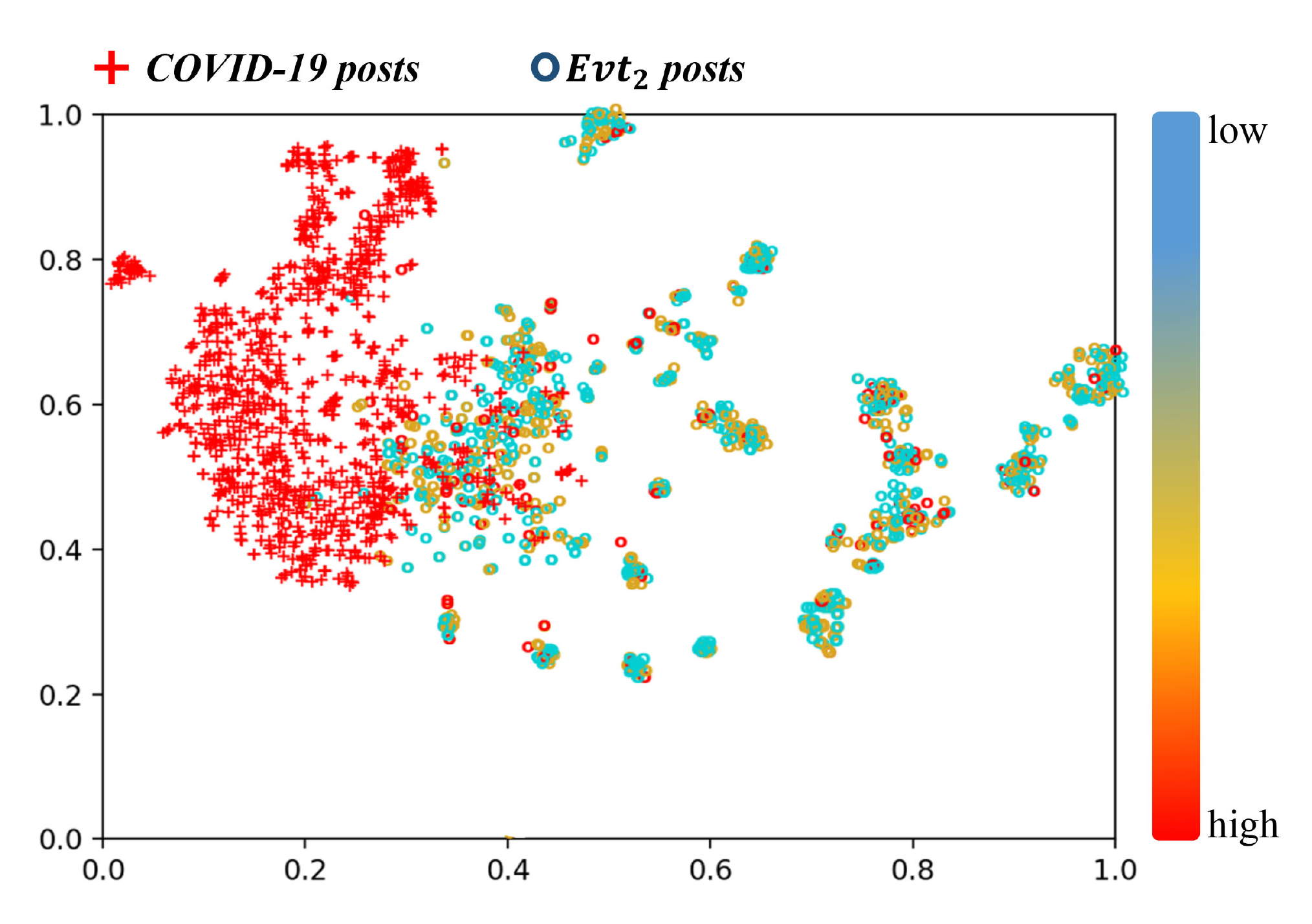}
\caption{The visualization of the learned weights of source samples in $Evt_{2}$. Specifically, the ``+'' means posts included in the unlabeled COVID-19 data in model training stage, while the ``o'' represents the posts from the $Evt_{2}$. The red circles represent the source event posts with higher weights, and blue circles represent the lower weights samples. In addition, the intermediate values determine their colors based on the color bar.} 
\label{weight_distribution}
\end{figure}

\subsubsection{Case studies for COVID-19 related fake news detection}
\begin{figure}
\includegraphics[width=\textwidth]{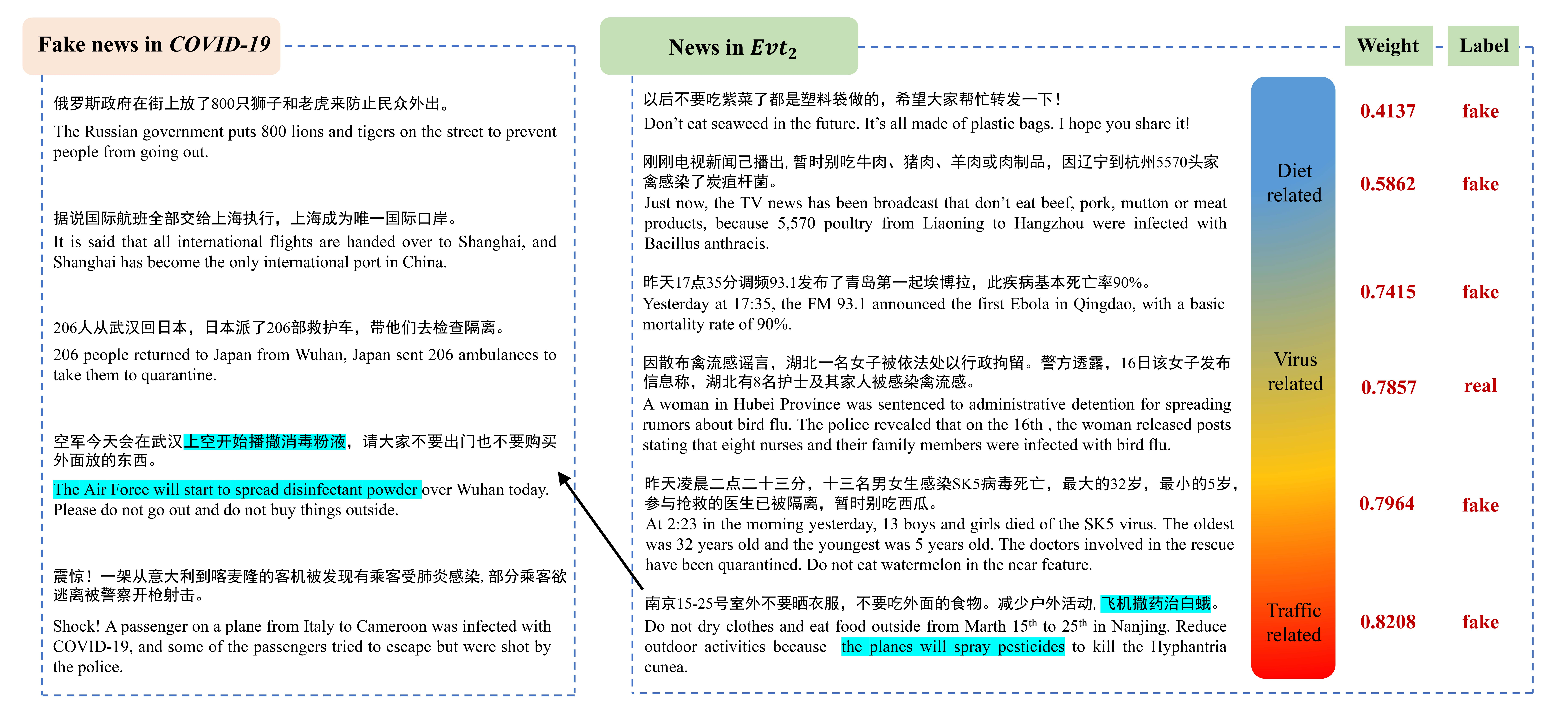}
\caption{The representative posts selected and ranked by \textit{MetaDetector}.} 
\label{case_study}
\end{figure}

We further analyze the validation of meta event knowledge transfer for fake news detection, and map the weight of each source event instance learned by \textit{MetaDetector} to the event data distribution through the gradient color. In this subsection, we take the task $Evt_{2}\rightarrow$ \textit{COVID-19} as an example. As shown in the Figure \ref{weight_distribution}, there are 1,000 sampling points from the data about COVID-19 and 1,500 sampling points from the $Evt_{2}$ in the training stage. The gradient colors indicate corresponding weight values (red represents the highest value to 1). It can be seen that most of the source posts, far away from target posts, are cyan (with lower weights), and that gold circles (with higher weights) are mostly distributed in the area overlapping with the target event. However, some circles that are farther away from the target red points are also given high weights, and vice versa, indicating that \textit{MetaDetector} not only pays attention to the transferable features of different events but also integrates the event-specific features. This ensures that our proposed model improves the generalization performance while maintaining the accuracy of target fake news detection. In addition, we select several representative news posts based on weight values to illustrate that the pseudo-event discrimination-based weighting mechanism cultivates the model to capture influential source posts for detecting target fake news, as shown in Figure \ref{case_study}. After abstracting the semantic information of sampled posts, we could observe that social posts related to \textit{diet} generally have lower weights, while \textit{viruses-related} posts have generally higher weights. This further reveals that directly utilizing traditional adversarial learning methods to align the source and target event data distributions can easily confuse the sub-semantic information underlying the whole event and cause the shared feature space to shift. We also notice that the sixth post of $Evt_{2}$ in Figure \ref{case_study} has a relatively high weight, probably because of a similar description with the forth post in COVID-19 (as highlighted in the Figure \ref{case_study}). 

\section{Conclusion and Future Work}
\label{Conclusion}
This work focuses on the upcoming fake news detection on social media, which is a practical problem ignored by the research community. The major challenge is how to learn the event-shared meta knowledge to alleviate the negative transfer. In order to tackle this problem, we propose a weighted adversarial event adaptation network based on unsupervised adversarial domain adaptation nets, namely \textit{MetaDetector}, which aims to extract transferable meta-knowledge for fake news. \textit{MetaDetector} utilizes the pseudo-event discrimination-based weighting mechanism to automatically recognize historical posts that are irrelevant or abnormal to the target events, which reduces the discrepancy of events and promotes the model generality to unseen events. Experiments on public datasets collected from Twitter and Sina Weibo demonstrate that the meta-knowledge is constructive to learning the representation space of upcoming events, and \textit{MetaDetector} outperforms the state-of-the-art fake news detection methods especially when the distribution shift between events is significant, which is constructive to alleviating the negative transfer.

For the future work, we will explore multi-source events knowledge transfer methods for detecting fake news in new events. The meta knowledge presented in this paper is currently from the perspective of transferable linguistic features. However, the transferability of information such as the \textit{logic}, \textit{causality}, and \textit{factual evidence} underlying different events is also worth exploring.

\begin{acks}
This work was partially supported by the National Science Fund for Distinguished Young Scholars (62025205, 61725205), National Key R\&D Program of China (2019QY0600), and the National Natural Science Foundation of China (No. 61960206008, 61902320, 61972319).
\end{acks}

\bibliographystyle{ACM-Reference-Format}
\bibliography{MetaDetector}


\end{document}